\documentclass[preprint,preprintnumbers,amsmath,amssymb]{revtex4}
\usepackage{graphicx}
\usepackage{epstopdf}
\usepackage{dcolumn}
\usepackage{bm}
\usepackage{color}

\begin{document}

\title{Spectroscopic and DFT studies of graphene intercalation systems on metals}

\author{Yu. S. Dedkov$^{1,2,}$\footnote{Corresponding author. E-mail: yuriy.dedkov@icloud.com} and E. N. Voloshina$^{3,}$\footnote{Corresponding author. E-mail: elena.voloshina@hu-berlin.de}}
\affiliation{$^1$\mbox{SPECS Surface Nano Analysis GmbH, Voltastra\ss e 5, 13355 Berlin, Germany}}
\affiliation{$^2$Fachbereich Physik, Universit\"at Konstanz, 78457 Konstanz, Germany}
\affiliation{$^3$Humboldt-Universit\"at zu Berlin, Institut f\"ur Chemie, 10099 Berlin, Germany}

\date{\today}

\begin{abstract}
Intercalation of different species under graphene on metals is an effective way to tailor electronic properties of these systems. Here we present the successful intercalation of metallic (Cu) and gaseous (oxygen) specimens underneath graphene on Ir(111) and Ru(0001), respectively, that allows to change the charge state of graphene as well as to modify drastically its electronic structure in the vicinity of the Fermi level. We employ ARPES and STS spectroscopic methods in combination with state-of-the-art DFT calculations in order to illustrate how the energy dispersion of graphene-derived states can be studied in the macro- and nm-scale experiments. 
\end{abstract}

\maketitle

\section{Introduction}

Intercalation compounds on the basis of different layered host materials attract a lot of attention in the last decades. The most common examples, which appeared naturally during studies of the transport properties of graphite, are the graphite intercalation compounds (GICs)~\cite{Dresselhaus:2002}, which can be formed in different ways via incorporation of atoms of metals or non-metals or big molecules between single graphite layers. Further studies lead to the discovery of intercalation compounds on the basis of layers of CoO$_2$ (Li$_x$CoO$_2$, Na$_x$CoO$_2$, etc.)~\cite{Bruce:1997kf}, transition metal dichalcogenides (Cu$_x$TiSe$_2$, Li$_x$MoS$_2$, etc.)~\cite{Chhowalla:2013fz} and many others. All these studies demonstrate that chemical and physical properties of the intercalant hosts might be modified via controllable tailoring their electronic band structure, sometimes leading to the observation of new phenomena, which were previously not observed for the parent material, like superconductivity in alkali-metal GICs~\cite{Weller:2005kt} or Cu$_x$TiSe$_2$~\cite{Morosan:2006hk}.

In case of intercalation compounds on the basis of graphite or transition metal dichalcogenides, inserting intercalant layers leads to the decoupling of 2D layers of the host material from each other allowing to perform studies of the electronic structure of the single layers, which is modified by the presence of intercalant. It is interesting that despite the 3D crystallographic structure of these materials and change of the periodicity in the system upon intercalation, the valence band structure of the resulting materials has a symmetry of the host material~\cite{Molodtsov:1998wk,Brandt:2001he,Molodtsov:2003jp}. This effect was nicely demonstrated in the angle-resolved photoelectron spectroscopy (ARPES) experiments on La-GIC compound~\cite{Molodtsov:1998wk,Molodtsov:2003jp}, where band structures of a 2D host graphene layers and the inserted La layer, which have different symmetries, give the photoemission signals in the Brillouine zone of the respective symmetry. It was shown that if the symmetry of the sub-systems is different and the interaction between atoms in the sub-unit is stronger than between units of the crystal, then the folded bands are not observed in the photoemission experiment. Similar consideration is also valid for the local spectroscopic experiments, such as scanning tunneling spectroscopy (STS) or transport measurements, where effect of additional periodicity can change the transport coefficients dramatically.

Graphene (gr), as a pure 2D material, is a subject of the intensive studies after its extraordinary transport properties, caused by the linear dispersion of the electronic states in the vicinity of the Fermi level ($E_F$), were discovered~\cite{Geim:2007a,CastroNeto:2009}. Besides the perspectives for graphene to be used in different technological applications, e.\,g. in gas sensors~\cite{Schedin:2007}, in touch screens~\cite{Ryu:2014fo}, or as a protective anticorrosion layer~\cite{Dedkov:2008d,Dedkov:2008e,Sutter:2010bx,Chen:2011a}, it is a perfect material to study different physical phenomena in 2D materials that in many cases leads to the observation of the new fascinating effects. For example, if exfoliated graphene flake is placed on the h-BN substrate, then the small missalignment of two sublattices leads to the formation of the moir\'e lattices in this system that influences the electronic spectrum of graphene and the so-called Hofstadter butterfly spectrum of electrons, which movement is influenced by simultaneously acting periodic potential and the external magnetic field~\cite{Hunt:2013ef,Ponomarenko:2013hl}.

Further example is the graphene/metal system~\cite{Tontegode:1991ts,Wintterlin:2009,Batzill:2012,Dedkov:2015kp}, where graphene is usually synthesised via chemical vapour deposition (CVD) technique. Here the interaction between graphene and the metallic substrate can be modified by placing different species (atoms of metals or non-metals, molecules, like CO, H$_2$O, or C$_{60}$) underneath graphene~\cite{Shikin:2000a,Dedkov:2001,Voloshina:2011NJP,Larciprete:2012aaa,Granas:2012cf,Granas:2013tl,Petrovic:2013vz,Voloshina:2016jd,Politano:2016jb}. The main goal of such studies is to decouple graphene from the metallic substrate, that in most cases changes the doping level of graphene (and even its sign) and to restore the original linear dispersion of the graphene-derived states in the vicinity of the Fermi level. Here ARPES and STS techniques allow to perform electronic structure studies on the macro- or/and nm-scale giving information about doping level, gap opening, band dispersion and renormalization, and also providing new physics, which shed light on the new phenomena in graphene~\cite{Dedkov:2015kp,Varykhalov:2008,Morgenstern:2011hh,Fedorov:2014aa,Dedkov:2015iza,Vita:2014aa}.

In the present work we demonstrate the application of space-integrated (ARPES) and local (STS) spectroscopic methods in combination with density-functional theory (DFT) calculations for investigation of the electronic band structure of the graphene-based intercalation systems on metals: gr/Cu/Ir(111) and gr/O/Ru(0001). Both systems show the electronic properties (doping level and band dispersions) which are different from those for the parent systems, demonstrating the effective ways to tailor physical and chemical properties of the graphene/metal interface. This work is an extended review of the previously published results~\cite{Voloshina:2016jd,Vita:2014aa}.

\section{Experimental details}

Graphene in both experiments was prepared on hot metallic substrates (preliminary cleaned by cycles of Ar$^+$-sputtering/annealing) via decomposition of C$_2$H$_4$ at a partial pressure of $p=1\times10^{-7}$\,mbar. Intercalation of Cu in gr/Ir(111) was performed via annealing of the Cu/gr/Ir(111) system with a nominal thickness of Cu-layer slightly more than $1$\,ML. The process of intercalation was monitored in the \textit{live}-XPS (x-ray photoelectron spectrosocpy) experiments. Oxygen intercalation in gr/Ru(0001) was performed with a stainless steel pipe, which end was placed in the close vicinity of the sample surface, at the relatively high gas pressure ($p_{O_2}=1.5\times10^{-4}$\,mbar) and sample temperature of $150^\circ$\,C. Homogeneity and cleanness of the systems before and after intercalation were controlled in the respective STM, XPS, and ARPES experiments. The more detailed description of the sample preparation procedures can be found elsewhere~\cite{Voloshina:2016jd,Vita:2014aa}.

All prepared samples were characterized at room temperature by means of STM/AFM performed with SPM Aarhus 150 equipped with KolibriSensor\texttrademark. In these measurements the sharp W-tip was used, which was cleaned \textit{in situ} via Ar$^+$ -sputtering. In the presented STM images the tunnelling bias voltage, $U_T$, is applied to the sample. Low-temperature STM measurements were performed in the SPECS JT-STM at the sample and tip temperature of $1$\,K. $dI/dV$ spectroscopy and mappings were performed at low temperatures using the lock-in-technique with a modulation voltage of $U_{mod} = 10$\,mV and a modulation frequency $f_{mod} = 684.7$\,Hz.

ARPES experiments were performed in the UHV station equipped with SPECS PHOIBOS\,150/2D-CCD analyzer and Ar/He UV-light source. The sample was placed on a 5-axis motorized manipulator, allowing for its precise alignment in the $k$ space. The sample was azimuthally pre-aligned in such a way that the polar scans were performed along the $\Gamma-\mathrm{K}$ direction of the graphene-derived BZ with the photoemission intensity on the channelplate images acquired along the direction perpendicular to $\Gamma-\mathrm{K}$. The final 3D data sets of the photoemission intensity as a function of kinetic energy and two emission angles were transformed in the respective data sets for the reciprocal space, $I(E_B, k_x, k_y)$ for the careful analysis, where $E_B$ is the binding energy of electrons and $k_{x,y}$ are two orthogonal components of the wave vector in BZ. Part of ARPES and XPS experiments on the intercalation process studies was performed at BESSY\,II (HZB Berlin).

The DFT calculations were carried out with the \texttt{VASP} program~\cite{Kresse:1994} using the projector augmented wave (PAW) method~\cite{Blochl:1994}, a plane wave basis set, and the generalised gradient approximation as parameterised by Perdew \textit{et al.}~\cite{Perdew:1996}. The plane wave kinetic energy cutoff  was set to $400$\,eV. The long-range van der Waals interactions were accounted for by means of the DFT-D2 approach~\cite{Grimme:2006}. The supercells used to model the graphene-substrate interfaces are constructed from a slab of $5$ metal layers with a graphene layer adsorbed from one side and a vacuum region of approximately $20$\,\AA. In the case of graphene/Ir(111) and graphene/Cu/Ir(111) this supercell has a $(9\times9)$ lateral periodicity with respect to metal and a $(10\times10)$ lateral periodicity with respect to graphene. In the case of graphene/O/Ru(0001) this supercell has a $(12\times12)$ lateral periodicity with respect to metal and a $(13\times13)$ lateral periodicity with respect to graphene. To avoid interactions between periodic images of the slab, a dipole correction is applied. During the structure relaxation, the positions of the carbon atoms as well as those of the top two layers of metal atoms (as well as oxygen atoms in the case of graphene/O/Ru(0001)) are allowed to relax. In the total energy calculations and during the structural relaxations the $k$-meshes for sampling the supercell Brillouin zone are chosen to be as dense as $6\times6$ and $3\times3$, respectively, and centred at the $\Gamma$-point. The band structures calculated for the studied systems were unfolded to the graphene $(1\times1)$ primitive unit cell according to the procedure described in Refs.~\cite{Popescu:2012bq,Medeiros:2014ka} using the \texttt{BandUP} code.

\section{Results and discussions}

\subsection{Systems preparation and characterization.}

\begin{figure}
\includegraphics[width=\linewidth]{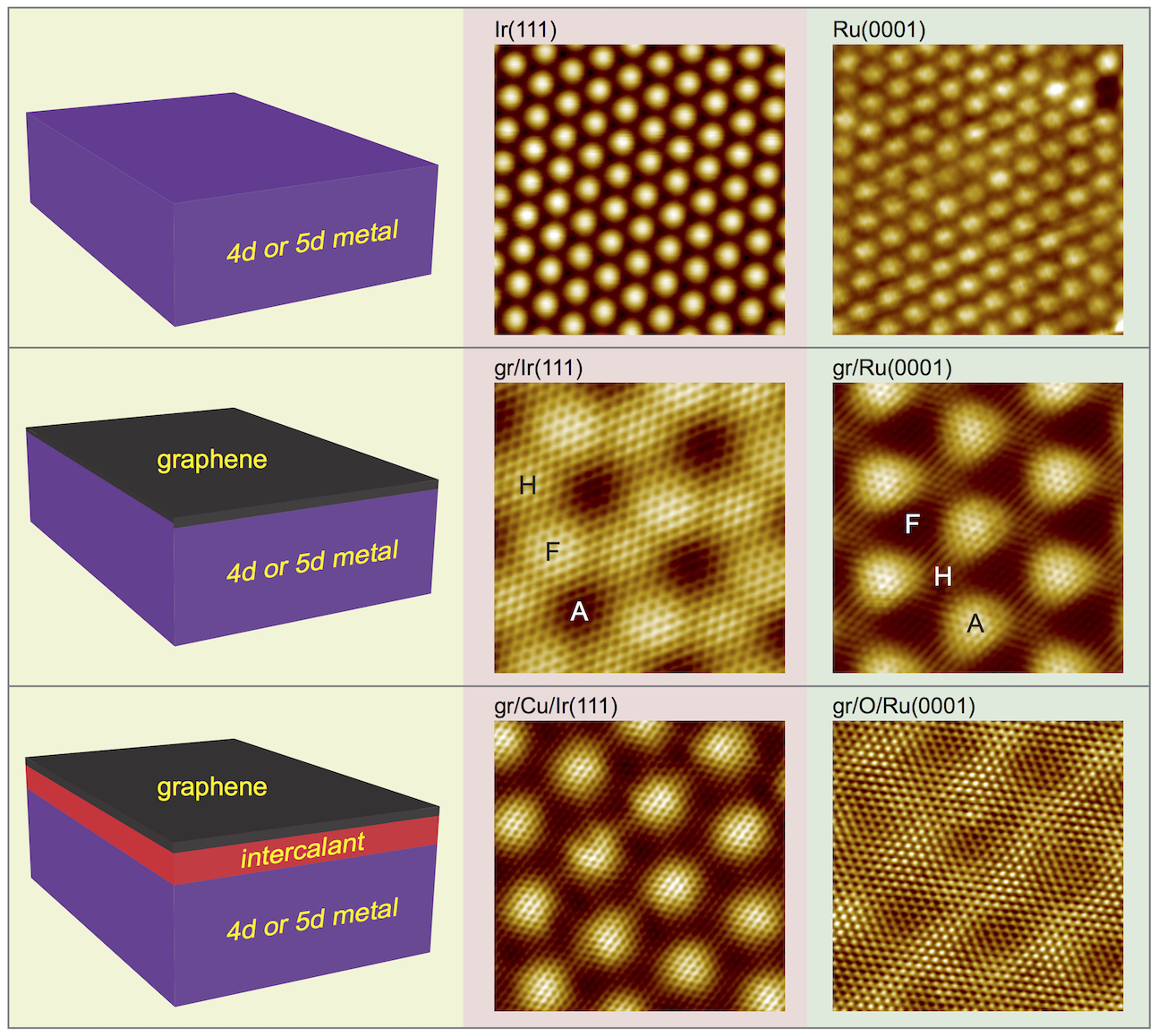}
\caption{Preparation steps (from top to bottom) of intercalation systems on the basis of graphene/metal. Top row: atomically resolved STM images of Ir(111) and Ru(0001); middle row: STM images of gr/Ir(111) and gr/Ru(0001) moir\'e structures; bottom row: STM images of gr/Cu/Ir(111) and gr/O/Ru(0001) intercalation systems.}
\label{Scheme_STM}
\end{figure}

Figure~\ref{Scheme_STM} summarizes the results of the samples preparation and characterization. Original, hexagonally packed clean surfaces of $4d$ or $5d$ metals, \textit{hcp} Ru(0001) or \textit{fcc} Ir(111) (top row), are used for the preparation of the high-quality graphene layers, which in these cases form the so-called moir\'e structures  with several nm periodicities (middle row). Here several high-symmetry adsorption positions for carbon atoms on the close-packed surfaces can be identified: ATOP, FCC, HCP (they are marked by the respective capital letters in Fig.~\ref{Scheme_STM}, middle row). They are determined with respect to the corresponding adsorption places of the metal surface, which are surrounded by the carbon ring of a graphene layer. For low bias voltages, graphene on Ir(111) is imaged in the \textit{inverted} contrast, when the topographically highest ATOP place of a graphene layer is imaged as a dark spot in the STM image as opposite to the HCP and FCC positions. Such an effect was explained by the formation of the corresponding interface states between graphene and Ir(111) as a result of an overlap between graphene $\pi$ and Ir\,$5d_{z^2}$ states~\cite{Voloshina:2013dq,Dedkov:2014di}. For graphene on Ru(0001) the \textit{direct} imaging contrast is prevailed in STM images for low bias voltages due to the domination of the topographic contribution in the STM imaging~\cite{Voloshina:2016jd,Stradi:2011be,Stradi:2012hw,Voloshina:2016kp}.

Intercalation of Cu and oxygen in gr/Ir(111) and gr/Ru(0001) (Figure~\ref{Scheme_STM}, bottom row), respectively, conserves the periodicity of the moir\'e lattices (pseudomorphic growth of Cu at the gr/Ir interface~\cite{Vita:2014aa} and $(2\times1)$-O structure at the gr/Ru interface~\cite{Voloshina:2016jd}), but leads to the dramatic changes in the STM imaging contrast. After intercalation the imaging contrast is \textit{direct} for gr/Cu/Ir(111) and the STM contrast for gr/O/Ru(0001) becomes extremely flat. Such changes in the imaging contrast for both systems are the reflection of the respective modifications in the electronic structure of graphene on the corresponding support and they are the subjects of the further ARPES and STS studies of the respective systems.

\begin{figure}
\includegraphics[width=\linewidth]{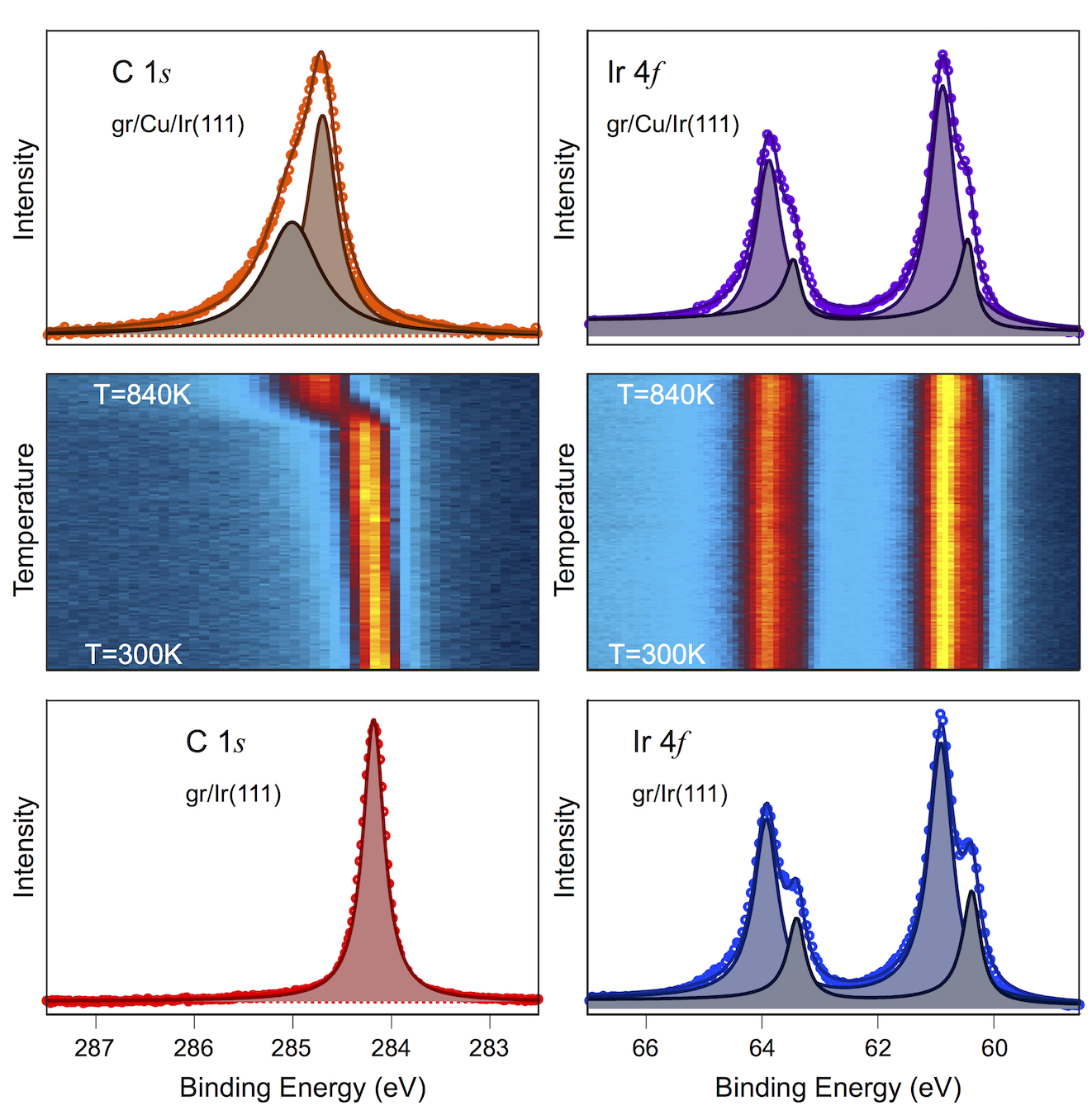}
\caption{XPS spectra of gr/Ir(111) (bottom) and gr/Cu/Ir(111) (top): C\,$1s$ (left column) and Ir\,$4f$ (right column). The respective fitting components are shown by the shaded areas. Middle row shows the evolution of the C\,$1s$ and Ir\,$4f$ intensity as a function of the annealing temperature resulting in formation of gr/Cu/Ir(111) at $840$\,K. }
\label{grCuIr_XPS}
\end{figure}

In order to control the formation of the gr/Cu/Ir(111) system, which STM images might be, at some imaging conditions, similar to those of gr/Ir(111), the process of Cu intercalation under graphene on Ir(111) was monitored in the \textit{live}-XPS measurements, where C\,$1s$ and Ir\,$4f$ core levels were acquired as functions of time with sample temperature ramped from room temperature to $850$\,K. Fig.~\ref{grCuIr_XPS} shows the high-resolved C\,$1s$ and Ir\,$4f$ XPS spectra taken before and after intercalation (bottom and top rows) as well as the respective photoemission intensity maps as a function of binding energy and sample temperature (middle row). From these intensity maps one can see that until $\approx820$\,K there are only gradual changes in the spectra: small shift of the C\,$1s$ line to the higher binding energy and the weak decrease of intensity of the interface component of the Ir\,$4f$ line (low binding energy shoulder). At $T\approx830$\,K drastic changes are observed: (i) C\,$1s$ line shifts stepwise to the higher binding energy by $\approx0.55$\,eV and (ii) the peak intensity of the ``interface'' components for the Ir\,$4f$ lines is reduced by $\approx30$\% (at the same time the peak intensity of the ``bulk'' components is reduced by $\approx18$\%). Both changes as well as the absence of any C\,$1s$ signal at the binding energies corresponding to gr/Ir(111) indicates the complete intercalation of the Cu layer and formation of the gr/Cu/Ir(111) intercalation system. Before and after Cu intercalation the C\,$1s$ XPS line can be fitted by one ($284.18$\,eV) and two ($284.69$\,eV and $285.01$\,eV) components, respectively, reflecting stronger interaction between graphene and Cu/Ir(111) with smaller gr/metal distance compared to gr/Ir(111), and that two distinct areas in the STM images are observed. For the Ir\,$4f$ spectra the energy splitting between ``bulk'' and ``interface'' components is reduced from $537$\,meV to $463$\,meV for systems before and after Cu intercalation, respectively. Intensity of the ``interface'' part is strongly reduced after Cu intercalation in gr/Ir(111).

\subsection{ARPES and DFT of gr/Ir(111) and gr/Cu/Ir(111)}

\begin{figure}
\includegraphics[width=\linewidth]{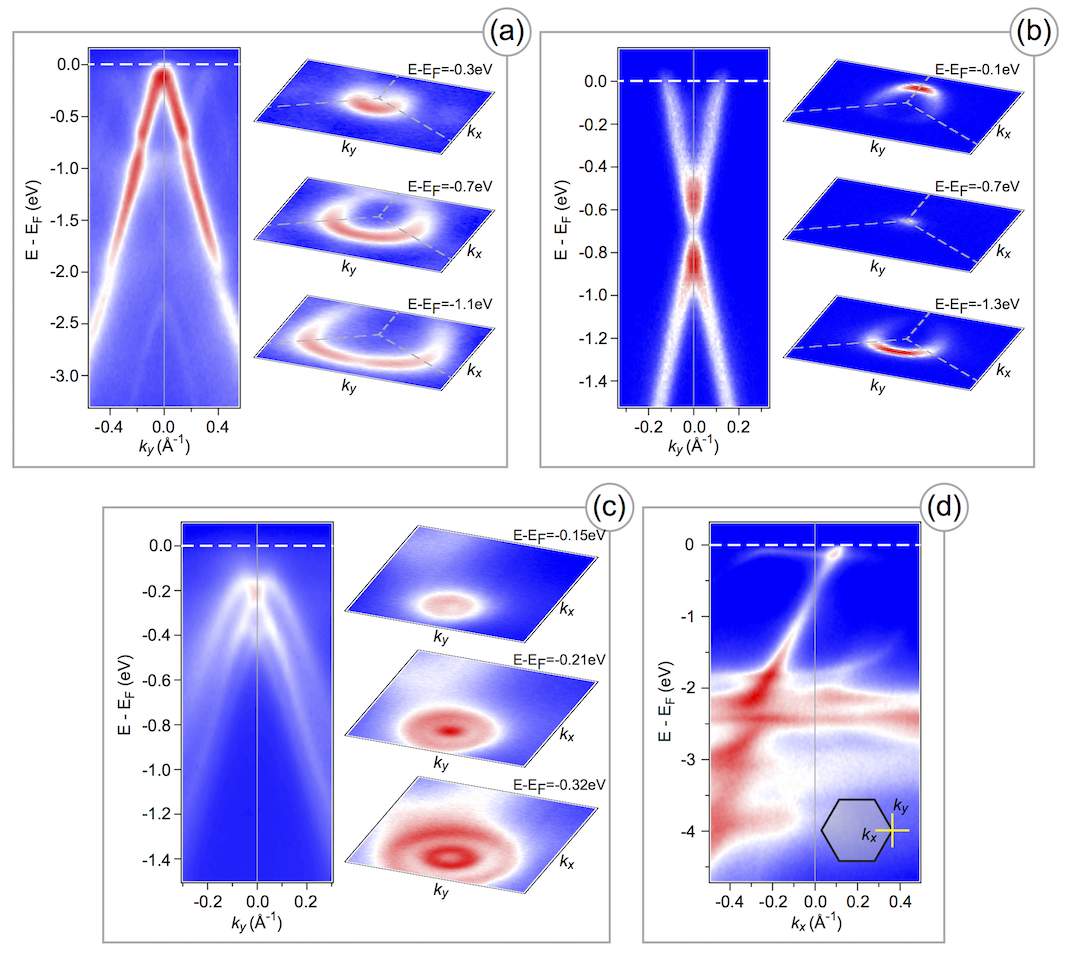}
\caption{ARPES intensity maps [$I(E_B,k_y)$] around the $\mathrm{K}$-point and the respective constant energy cuts [$I(k_x,k_y)$] for (a) gr/Ir(111) and (b) gr/Cu/Ir(111). Data were obtained with photon energy of $h\nu=40.8$\,eV. (c) ARPES intensity map [$I(E_B,k_y)$] around the $\Gamma$-point and the respective constant energy cuts [$I(k_x,k_y)$] showing interface state in gr/Ir(111). Data were acquired with non-monochromatized Ar\,I line (Ar\,I$\alpha$/Ar\,I$\beta$, $h\nu=11.83$\,eV/$11.62$\,eV). (d) APRES intensity map along $\Gamma-\mathrm{K}$ for gr/Cu/Ir(111) obtained with photon energy $h\nu=65$\,eV. Inset shows the sketch of the hexagonal Brillouine zone with marked $k_x, k_y$ directions.}
\label{grCuIr_ARPES}
\end{figure}

Figure~\ref{grCuIr_ARPES} compiles the results on the electronic structure studies of graphene on Ir(111) and Cu/Ir(111). ARPES intensity map for gr/Ir(111) shows a clear dispersion of the graphene-derived $\pi$ band along the $k$ direction perpendicular to $\Gamma-\mathrm{K}$ in the Brillouine zone [Fig.~\ref{grCuIr_ARPES}(a)]. According to the recent theoretical works~\cite{Vita:2014aa,Voloshina:2013dq,Busse:2011}, the interaction between graphene and Ir(111) is mainly governed by the van der Waals forces and the \textit{hybridization} between graphene-derived and Ir valence band states is relatively small. This leads to the observation of the linear dispersion of the graphene $\pi$ states around $E_F$, with the position of the Dirac point at approximately $100\pm10$\,meV above $E_F$ (graphene is $p$-doped). The influence of the Ir(111) substrate is manifested via observation of the so-called replica bands in the ARPES intensity map and opening of the mini-gaps at $E-E_F\approx-0.76$\,eV and $E-E_F\approx-2.1$\,eV according to the \textit{avoid-crossing} mechanism~\cite{Dedkov:2014di,Pletikosic:2009,Rusponi:2010}. These effects are due to the additional moir\'e lattice periodicity (of $\approx25$\,\AA) observed for this system.

The relatively weak interaction between graphene and Ir(111) does not destroy the original Rashba-split surface state observed on Ir(111) around the $\Gamma$ point [Fig.~\ref{grCuIr_ARPES}(c); here data were acquired with non-monochromatized Ar\,I line (Ar\,I$\alpha$/Ar\,I$\beta$, $h\nu=11.83$\,eV/$11.62$\,eV), leading to two sets of parabolas separated by energy of $0.21$\,eV]. Adsorption of graphene on Ir(111) leads only to the upward energy shift of the surface state from $E-E_F=-0.34$\,eV at $\Gamma$ for Ir(111) to $E-E_F=-0.21$\,eV at $\Gamma$ for gr/Ir(111), that can be assigned to the stronger localization of the surface state wave function upon graphene adsorption. This behavior is similar to the one observed for the same system~\cite{Varykhalov:2012ec} as well as for Au(111) and Ag(111) covered by graphene~\cite{Leicht:2014jy,Tesch:2016bd}.

Intercalation of Cu in gr/Ir(111) leads to the significant changes in the observed ARPES picture [Fig.~\ref{grCuIr_ARPES}(b)]. For the formed intercalation system, the energy shift of the Dirac point is observed, resulting in the $n$-doping of graphene, $E_D-E_F=-0.69\pm0.01$\,eV. The most important observation is the opening of the energy gap of $(0.36\pm0.01)$\,eV for the graphene $\pi$ states directly at the Dirac point. Similar behavior was also observed for the other intercalation systems with Cu, Au, and Ag used as intercalants~\cite{Enderlein:2010,Varykhalov:2010a,Papagno:2013ew}. It is interesting to note that a substantial \textit{hybridization} between graphene $\pi$ state and Cu\,$3d$ states is observed in the binding energy range of $\approx2-4$\,eV. This effect is reflected in the opening of the energy gaps at the energy and wave-vector values, where $\pi$ states are intersected by Cu\,$3d$ bands. As will be shown later this effect of \textit{hybridization} and space-overlap of the graphene-derived $\pi$ states with the Cu\,$3d$ states of different symmetry has an important implication on the spectral function of graphene $\pi$ states at the Dirac point.

\begin{figure}
\includegraphics[width=0.65\linewidth]{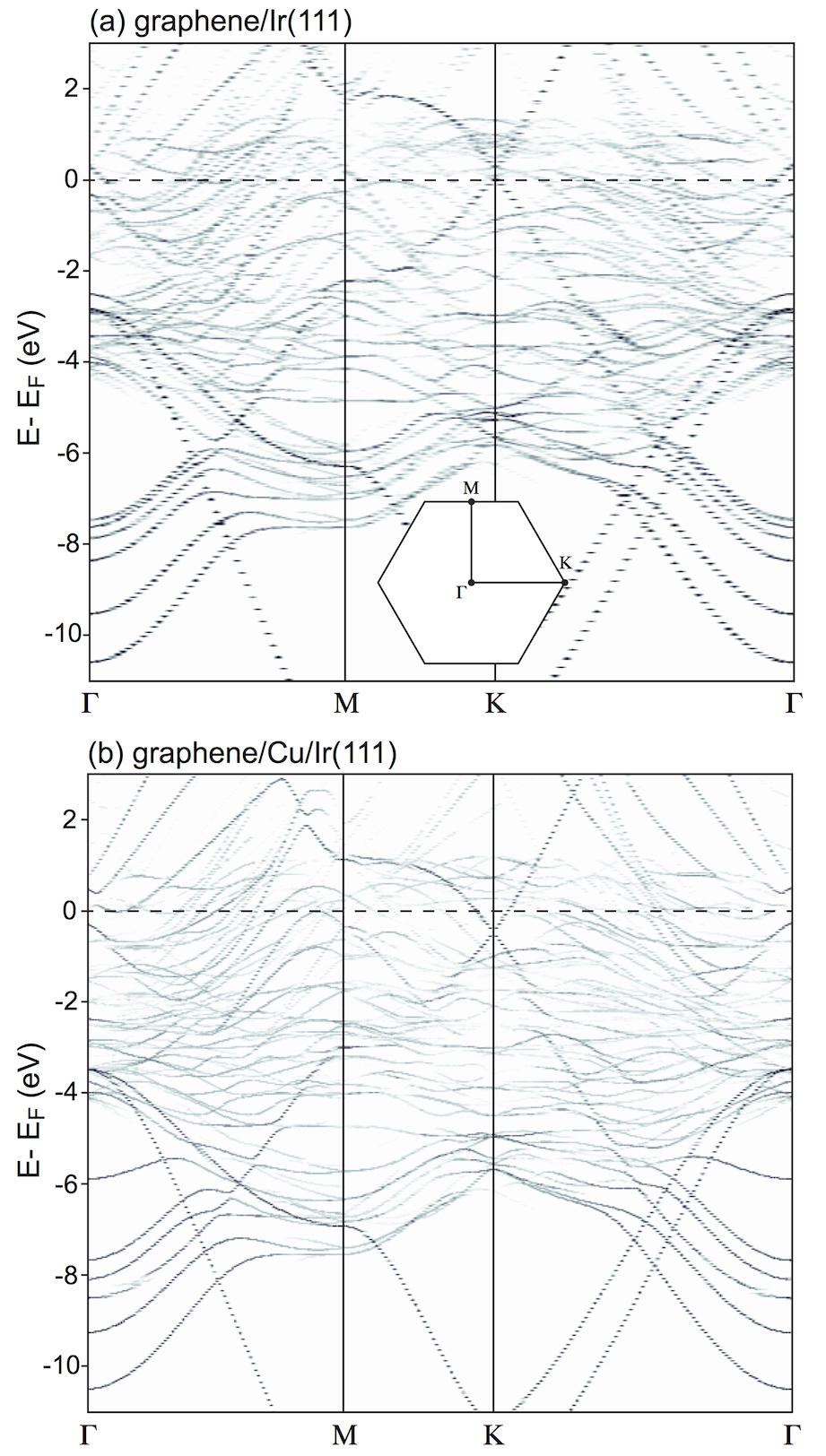}
\caption{Unfolded to the primitive $(1\times1)$ graphene unit cell the band structures of a graphene layer on (a) Ir(111) and (b) Cu/Ir(111).}
\label{grCuIr_theory}
\end{figure}

Figure~\ref{grCuIr_theory} shows the calculated band structure of a graphene layer on Ir(111) and Cu/Ir(111) for the corresponding supercell and then unfolded to the primitive $(1\times1)$ unit cell of graphene (band structures are presented along the respective high-symmetry directions of the graphene Brillouine zone shown as an inset in panel (a)). In both panels the dispersion of the graphene-derived $\pi$ states in the vicinity of the Fermi level can be easily distinguished. For gr/Ir(111) [Fig.~\ref{grCuIr_theory}(a)] a clear $p$-doping of a graphene layer is detected with the calculated position of the Dirac point of $E_D-E_F=0.11$\,eV. Closer analysis of the region around the Dirac point shows the existence of the energy gap of $0.22$\,eV in the energy dispersion of the $\pi$ band. Such gap with the width of $\approx100$\,meV was previously observed in the ARPES experiments on the electron-doping of gr/Ir(111) via K-adsorption~\cite{Starodub:2011a}. The appearance of this gap might be explained by the effect of \textit{hybridization} between graphene states of the $p_z$ symmetry ($\pi$ band) and the Ir\,$d_{z^2}$ surface state localized around the $\mathrm{K}$-point, which was recently detected in the experiment~\cite{Pervan:2015cq}.

In case of the gr/Cu/Ir(111) intercalation system, a graphene layer has an $n$-doping with a position of the Dirac point at $E_D-E_F=-0.49$\,eV. At the same time the energy gap of $0.11$\,eV is opened directly at the Dirac point. Taking into account the size of the studied system, the agreement between experimental and theoretical values of doping and the width of the band gap is rather good. There is also a series of the energy gaps in the energy range $E-E_F=-1.5...-5$\,eV, which appear at energies and $k$-vector values where graphene $\pi$ band is crossed by the Cu\,$3d$ bands leading to the formation of the \textit{hybrid} states. The deep analysis performed in Refs.~\cite{Vita:2014aa,Voloshina:2014jl} shows that, in case of graphene on the close-packed surface of a $d$ metal, $p_z$ orbitals of two carbon atoms in the graphene unit cell overlap with $d$ orbitals of the underlying metal atom, which have different symmetry. In general case~\cite{Voloshina:2014jl} we can say that the following \textit{hybrid} states are formed: $p_z(C^{top})+d_{z^2}$ and $p_z(C^{fcc})+d_{xz,yz}$, where $C^{top}$ and $C^{fcc}$ are carbon atoms in the unit cell occupying $top$ and $fcc$ high-symmetry adsorption positions above the close-packed surface. In case of the free-standing graphene, where carbon atoms are identical, the electronic states of two carbon sublattices are degenerate at the Dirac point that leads to the zero density of states at this point. If we consider graphene on $d$-metal, then symmetry between two carbon sublattices is broken and electronic states in the vicinity of the Dirac point have different symmetry as discussed above. This effect leads to the lifting of the degeneracy and opening of the energy gap directly at the Dirac point, which width is determined by the hybridization strength between graphene $\pi$ and metal $d$ states~\cite{Vita:2014aa,Voloshina:2014jl}.

\subsection{STS and DFT of gr/O/Ru(0001)}

\begin{figure}
\includegraphics[width=0.65\linewidth]{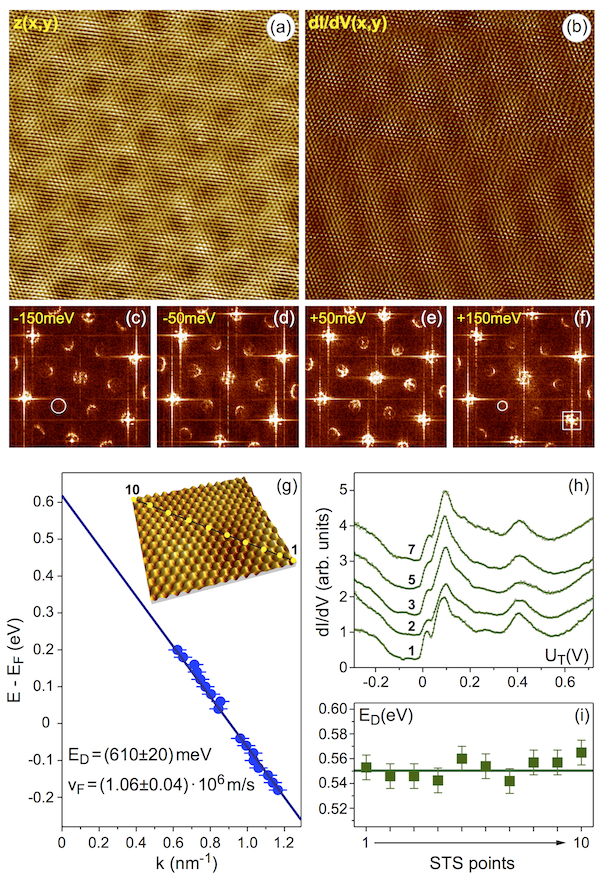}
\caption{Combined STM/STS measurements on gr/O/Ru(0001): (a) $z(x,y)$ and (b) $dI/dV(x,y)$. Scanning parameters: $20\times20\,\mathrm{nm}^2$, $U_T=+50$\,mV, $I_T=200$\,pA, $f_{mod}=684.7$\,Hz, $U_{mod}=10$\,mV. (c-f) Series of the FFT images obtained after transformation of the respective $dI/dV$ maps acquired at $U_T$ marked in every image. White circles mark the intervalley scattering features used in the analysis. (g) Extracted from the FFT maps energy dispersion of the carriers in graphene on O/Ru(0001): solid circles -- extracted points, solid line -- the respective linear fit with $E_D$ and $v_F$ marked in the figure. (h) Single $dI/dV$ spectra acquired along the line marked in the STM image shown as an inset in panel (g). (i) Extracted from $dI/dV$ spectra in (h) the position of $E_D$.}
\label{grORu_STS}
\end{figure}

The crystallographic structure of gr/O/Ru(0001) was studied in details by means of STM and non-contact atomic force microscopy (AFM) in Ref.~\cite{Voloshina:2016jd} and it was shown that graphene in this system is extremely flat with the corrugation of the moir\'e structure of only $0.1$\,\AA. The electronic structure of this nearly free-standing graphene was studied by means of STS on the local scale. The results of these experiments are compiled in Fig.~\ref{grORu_STS}.

The STM image of the gr/O/Ru(0001) system [Fig.~\ref{grORu_STS}(a)] shows drastic changes in the morphology of the system compared to the strongly corrugated parent gr/Ru(0001)~\cite{Voloshina:2016jd,Marchini:2007,Iannuzzi:2013fe} [Fig.~\ref{Scheme_STM}]. Intercalation of oxygen in gr/Ru(0001) leads to the significant reduction of the corrugation of the graphene moir\'e lattice from $1.27$\,\AA\ for the parent system~\cite{Voloshina:2016kp} to $0.12$\,\AA\ for gr/O/Ru(0001)~\cite{Voloshina:2016jd}. This effect together with the previously observed by ARPES strong $p$-doping of graphene in the later system~\cite{Sutter:2010a} indicates the decoupling of graphene from the substrate and restoring of the nearly free-standing character of the electronic states of graphene.

In order to obtain information about electronic structure of graphene in the intercalation system we performed local scale STS experiments on gr/O/Ru(0001). In the first approach we collected a series of $dI/dV$ maps at different bias voltages ($U_T$). One of such map acquired at $U_T=+50$\,mV is shown in Fig.~\ref{grORu_STS}(b). Different mechanisms for the scattering of the electron waves in graphene lead to the formation of the interference picture, which can be identified in such $dI/dV$ map together with the signals originating from the atomic graphene and moir\'e lattices. Fast Fourier Transformation (FFT) analysis of such $dI/dV$ maps [Fig.~\ref{grORu_STS}(c-f)] allows to extract the energy dispersion of the carriers involved in the scattering processes. In the FFT image several spots can be identified. First set, marked by the white rectangle in Fig.~\ref{grORu_STS}(f), is due to the reciprocal lattice of graphene (here the main central spot is surrounded by six spots originating from the moir\'e lattice of the system). The second set of spots in the FFT image, marked by the white circle, is placed at the positions of the $(\sqrt{3}\times\sqrt{3})R30^\circ$ structure of the reciprocal lattice, i.\,e. they are centred around the $\mathrm{K}$-points of the graphene Brillouine zone. This ring-shape structure is formed by the intervalley scattering of electrons between two equivalent points in BZ, $\mathrm{K}$ and $\mathrm{K'}$. The radius of this structure $2k$ can be used for the calculation of the wave vector of the scattering electron wave at the energy $E=eU_T$~\cite{Voloshina:2016jd,Leicht:2014jy,Tesch:2016bd,Simon:2011dv,Mallet:2012ib,Jolie:2014ev}. Such analysis performed for the series of $dI/dV$ images collected at different $U_T$ allows to extract the energy dispersion of the carriers in graphene around the $\mathrm{K}$-point of BZ. The results are shown in Fig.~\ref{grORu_STS}(g) by solid circles. Linear fit of these data leads to $E_D=(610\pm20)$\,meV and $v_F=(1.06\pm0.04)\cdot10^6$\,m/s. The extracted value of $E_D$ was compared with the one obtained from the single $dI/dV$ curves collected along the line in the STM image [inset of Fig.~\ref{grORu_STS}(g)]. The intensity dip around $550$\,meV [Fig.~\ref{grORu_STS}(h,i)] was assigned to the position of $E_D$ for graphene on O/Ru(0001) and this value is in rather good agreement with the one obtained in the previous analysis.

Experimental STM/STS results for gr/O/Ru(0001) were analysed in the framework of DFT. The parent gr/Ru(0001) system is modelled in the slab geometry, where a graphene layer with a $(13\times13)$ periodicity is placed on a Ru(0001) surface with a $(12\times12)$ periodicity, thus, forming the moir\'e lattice. Such structure was found to be adequate for the description of this system, giving the accurate description of the crystallographic and electronic structure of gr/Ru(0001)~\cite{Voloshina:2016jd,Voloshina:2016kp}. In our analysis we use the structure where after intercalation of O$_2$ in gr/Ru(0001), the oxygen atoms form the $p(2\times1)$ structure with $0.5$\,ML coverage with respect to Ru(0001) at the interface between graphene and Ru. It is supported by the experimental observations as well as by the calculated doping level~\cite{Voloshina:2016jd,Sutter:2010a,Sutter:2013kw,Dong:2015ig}. 

\begin{figure*}
\includegraphics[width=\linewidth]{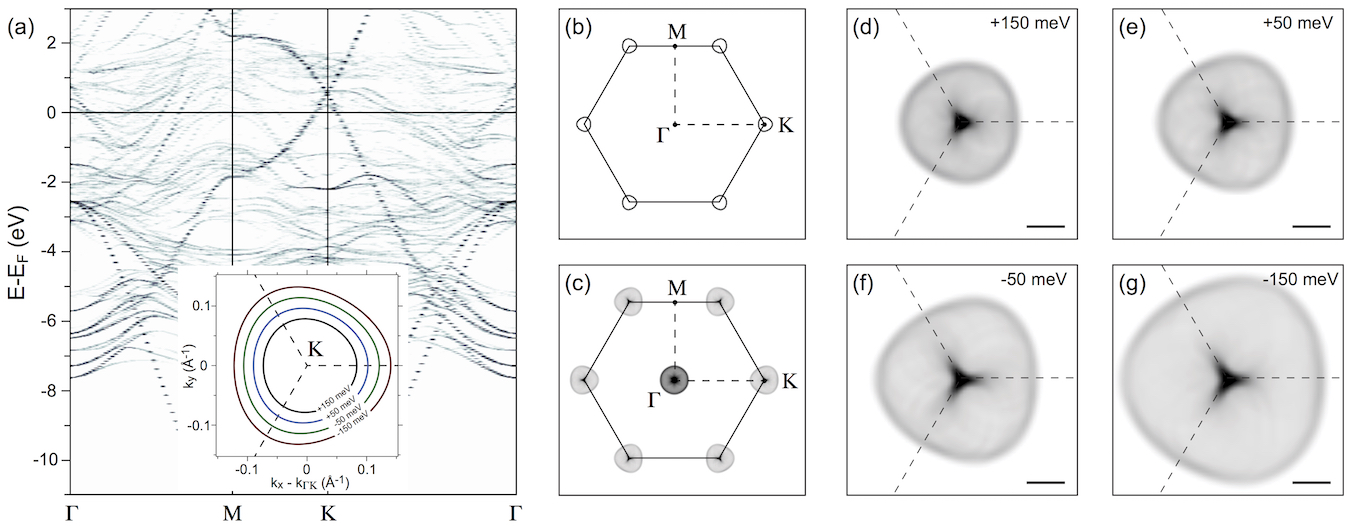}
\caption{(a) Band structure of gr/O/Ru(0001) along the $\Gamma-\mathrm{M}-\mathrm{K}-\Gamma$ direction of BZ. Inset shows TB-calculated CECs around the $\mathrm{K}$-point corresponding to the experimental values of $U_T$ (marked for every curve) and the Dirac point of $E_D-E_F=+600$\,meV. (b,c) CEC and the calculated FFT-STS map for energy $E-E_F=-750$\,meV for free-standing graphene, corresponding to $U_T=-150$\,meV used in the experiment. (d-g) Series of the calculated FFT-STS structures around the $\mathrm{K}$-point corresponding to the experimental values of $U_T$ marked in every image. Scale bar is $1\,\mathrm{nm}^{-1}$.}
\label{grORu_theory}
\end{figure*}

Figure~\ref{grORu_theory}(a) shows the calculated band structure of the graphene/O/Ru(0001) system in the supercell geometry described earlier and unfolded for the $(1\times1)$ unit cell of graphene. Although the complete band structure has a ``spaghetti-like'' view due to the folding of the bands, the main graphene-derived bands, $\pi$ and $\sigma$, can be easily recognised. One can clearly see that after intercalation of oxygen in gr/Ru(0001), graphene becomes nearly free-standing and its electronic states are completely decoupled from the Ru substrate (compared to gr/Ru(0001)~\cite{Voloshina:2016jd,Sutter:2010a,Brugger:2009}). Analysis of the electronic structure in the vicinity of the $\mathrm{K}$-point of graphene in this system shows that the $\pi$ band has a linear dispersion with a position of the Dirac point $E_D-E_F=+536$\,meV and $v_F=1.01\cdot10^6$\,m/s, indicating the strong $p$-doping of graphene. These values are in rather good agreement with experimental data presented above.

In order to demonstrate the effects observed in the STM/STS experiments we use the position of the Dirac point of $E_D-E_F=+600$\,meV, which is very close to the one obtained in the experiment (the difference of $10$\,meV does not dramatically alter the obtained results). The band structure of graphene in the vicinity of the Dirac point in the tight-binding (TB) approach can be expressed as $E_{\pm}=\pm t\sqrt{3+f(\mathbf{k})}$, where $f(\mathbf{k})=2\cos(\sqrt{3}k_ya)+4\cos(\sqrt{3}/2k_ya)\cos(3/2k_xa)$~\cite{CastroNeto:2009} ($t\approx2.8$\,eV is the nearest-neighbour hopping energy; $a=1.42$\,\AA\ is the distance between carbon atoms; $k_{x,y}$ are components of the wave vector). FFT maps presented in Fig.~\ref{grORu_STS}(c-f) were obtained at bias voltages $\pm50$\,meV and $\pm150$\,meV, corresponding to $E-E_F=-450, -550, -650, -750$\,meV for free-standing graphene. The respective constant energy cuts (CECs) for the $p$-doped graphene around the $\mathrm{K}$-point are shown as an inset of Fig.~\ref{grORu_theory}(a) with the corresponding $U_T$ values marked in the figure. From these data sets the clear trigonal warping of the $\pi$ bands with increasing of $E$ is visible that can be compared with experimental data.

Calculated CECs at different energies with respect to $E_D$ (and respectively at different $U_T$) can be used for modelling of the FFT scattering maps in STS experiments. As was shown in Refs.~\cite{Simon:2011dv,Simon:2007fd}, the power spectrum in the FFT map can be build on the basis of a joint-density-of-states (JDOS) approach, where one simply counts the number of pairs of $(\overrightarrow{k},\overrightarrow{k'})$, giving a scattering vector $\overrightarrow{q}$ with the same length and direction. For simplicity one can consider such calculations as a self-correlation procedure for the particular CEC image taken at the certain energy. Fig.~\ref{grORu_theory} (b) and (c) show the calculated CEC and FFT-STS maps obtained in the JDOS approach for the free-standing graphene corresponding to the energy of $E-E_F=-750$\,meV (corresponds to the value of the bias voltage of $U_T=-150$\,meV in the STS experiments on gr/O/Ru(0001)), where \textit{ring-like} structure around the $\mathrm{K}$-points is assigned to the intervalley scattering of the electron waves between equivalent $\mathrm{K}$ and $\mathrm{K'}$. The FFT maps obtained in the experiment shown in Fig.~\ref{grORu_STS}(c-f) can be compared with the respective theoretical results shown in Fig.~\ref{grORu_theory}(d-g) and good agreement between both sets of data is found.

\section{Conclusions}

In the present manuscript we demonstrate the successful application of the space-resolved (local) and space-integrated spectroscopic methods for the investigation of the electronic structure of the graphene intercalation systems formed on metals. All experimental data were reproduced in the DFT calculations allowing to understand the mechanisms leading to the formation of the spectrum of the charge carriers in graphene (doping level, gap openings, etc.). Two cases, of metal (Cu) and non-metal (oxygen) intercalation, were considered. In both cases the reversing of the doping level of graphene compared to the parent systems was observed. In the ARPES studies of gr/Ir(111) and gr/Cu/Ir(111) we followed the changes in the electronic structure and found $n$-doped graphene in the later case, with the position of the Dirac point of $E_D-E_F=-0.69$\,eV and the band gap of $0.36$\,eV directly at $E_D$. Symmetry analysis of the respective states around $E_D$ allows to draw the mechanism, which is responsible for the modifications of the band dispersion around the Dirac point. After intercalation of oxygen in gr/Ru(0001), a strong $p$-doping of graphene was found, which was confirmed in our DFT calculations. Band dispersion in the vicinity of $E_F$ of the graphene $\pi$ states for the obtained intercalation system was extracted in the analysis of the FFT-STS maps and the main results were reproduced in the framework of the JDOS approach for the strongly $p$-doped nearly free-standing graphene.

\section*{Acknowledgements} The authors thank H. Vita, S. B\"ottcher, K. Horn, T. Kampen, A. Thissen, N. Berdunov, M. Fonin, S. Lizzit, and R. Larciprete for useful discussions.  The High Performance Computing Network of Northern Germany (HLRN-III) is acknowledged for computer time. Financial support from the German Research Foundation (DFG) through the grant VO1711/3-1 within the Priority Programme 1459 ``Graphene'' is appreciated.

\clearpage

\clearpage

\clearpage

\clearpage

\clearpage

\clearpage

\clearpage

\end{document}